\documentclass{MIPRO}

\usepackage{cite}
\usepackage{amsmath,amssymb,amsfonts}
\usepackage{algorithmic}
\usepackage{graphicx}
\usepackage{textcomp}
\usepackage{xcolor}
\usepackage[T1]{fontenc}
\usepackage{multirow}
\usepackage{multicol}
\usepackage{array}
\usepackage{url}
\usepackage{dblfloatfix}

\usepackage{physics}
\usepackage{hyperref}
\hypersetup{
	colorlinks = true,
	linkcolor = blue,
	linkbordercolor = {white}
}

\DeclareMathOperator{\avg}{avg}
\newcommand{\refine}{\Gamma_\mathrm{r}}
\newcommand{\derefine}{\Gamma_\mathrm{d}}
\renewcommand{\vec}[1]{\boldsymbol{#1}}
\renewcommand{\L}{\mathcal{L}}

\usepackage[dvipsnames]{xcolor}

\graphicspath{{}}

\begin{document}
	
\title{Meshless $h$-adaptive Solution for non-Newtonian Natural Convection in a Differentially Heated Cavity}

\author{
	\IEEEauthorblockN{Miha 
		Rot\IEEEauthorrefmark{1}\IEEEauthorrefmark{2},
		Gregor Kosec\IEEEauthorrefmark{2}}
	
	\IEEEauthorblockA{\IEEEauthorrefmark{1}Jozef Stefan International 
		Postgraduate School, Jamova cesta 39, 1000 Ljubljana, Slovenia}
	\IEEEauthorblockA{\IEEEauthorrefmark{2}``Jožef Stefan'' Institute, 
		Parallel and Distributed Systems Laboratory, Jamova cesta 39, 1000 
		Ljubljana, Slovenia\\
		miha.rot@ijs.si}
	
	\thanks{The authors would like to acknowledge the financial support of the Slovenian Research and Innovation Agency (ARIS) research core funding No.\ P2-0095, Young Researcher programme PR-10468, and project funding N2-0275. Funded by National Science Centre, Poland under the OPUS call in the Weave programme 2021/43/I/ST3/00228. This research was funded in whole or in part by National Science Centre (2021/43/I/ST3/00228). For the purpose of Open Access, the author has applied a CC-BY public copyright licence to any Author Accepted Manuscript (AAM) version arising from this submission.}
}

\maketitle

\begin{abstract}
	One of the main challenges in numerically solving partial differential equations is finding a discretisation for the computational domain that balances the accurate representation of the underlying field with computational efficiency. Meshless methods approximate differential operators based on the values of the field in computational nodes, offering a natural approach to adaptivity. The density of computational nodes can either be increased to enhance accuracy or decreased to reduce the number of numerical operations, depending on the properties of the intermediate solution.
	In this paper, we utilise an adaptive discretisation approach for the numerical simulation of natural convection in non-Newtonian fluid flow. The shear-thinning behaviour is interesting both due to its numerous occurrences in nature, blood being a prime example, and due to its properties, as the decreasing viscosity with increasing shear rate results in sharper flow structures. We focus on the de Vahl Davis test case, a natural convection driven flow in a differentially heated rectangular cavity. The thin boundary layer flow along the vertical boundaries makes this an ideal test case for refinement. We demonstrate that adaptively refining the node density enhances computational efficiency and examine how the parameters for adaptive refinement affect the solution.
\end{abstract}

\renewcommand\IEEEkeywordsname{Keywords}

\section{Introduction}

\begin{figure*}[b]
	\includegraphics[width=\linewidth]{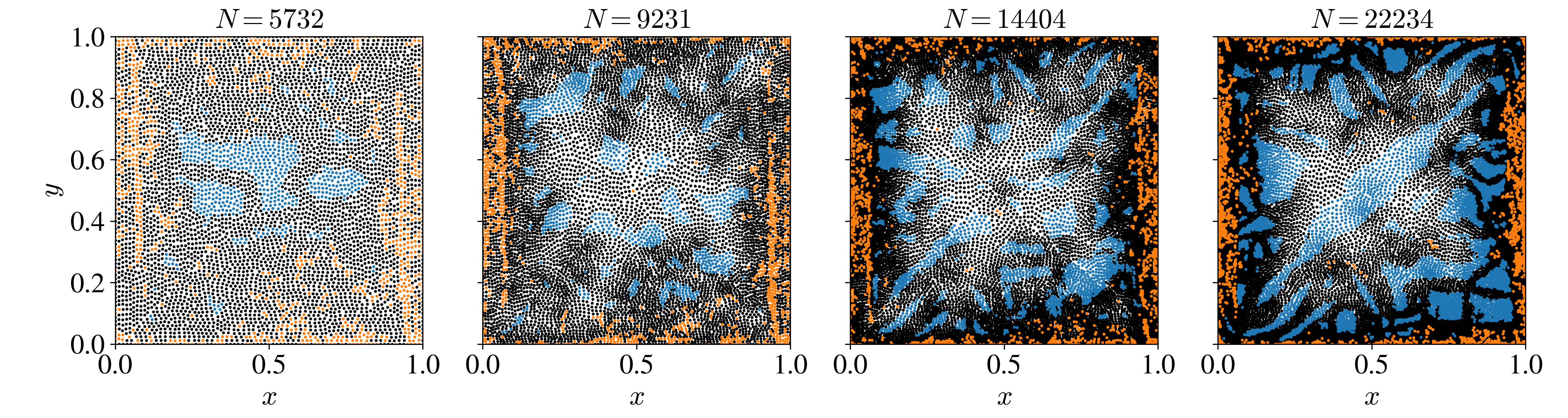}
	\caption{Node distributions during the first four adaptivity steps for the de Vahl Davis case. Nodes marked for refinement are plotted in orange, nodes marked for derefinement in blue, and neutral nodes in black.}
	\label{fig:refineDerefineScatter}
\end{figure*}

Adaptive discretisation is a well-established strategy for improving the
efficiency of numerical solutions of partial differential equations. The
core idea is to concentrate computational effort in regions where the
solution exhibits complex behaviour, such as steep gradients or boundary
layers, while keeping the discretisation coarse elsewhere. In the context of
mesh-based methods, $h$-adaptivity (inserting or removing elements),
$p$-adaptivity (varying the polynomial order), and $r$-adaptivity
(relocating nodes) are mature and extensively
studied~\cite{ZienkiewiczZhu1987,Segeth2010,Mitchell2014}.

Meshless methods, which approximate differential operators on scattered
computational nodes without an underlying connectivity structure, are
naturally suited to adaptive discretisation because the node density can be
modified without the topological constraints imposed by a mesh. Localised
RBF meshless methods have been successfully applied to fluid flow and
conjugate heat transfer problems~\cite{DivoKassab2007, Zamolo2019}, while
adaptive meshless procedures have been explored in various contexts, including linear elasticity~\cite{Slak2019, Jancic2025hp}, dendrite growth~\cite{dobravec2026adaptiveGFDM}, natural convection driven flow~\cite{Bacer2025}, and operators themselves \cite{Reeger2024}.

In our recent work~\cite{Rot2025} the natural
convection of a non-Newtonian power-law fluid in a differentially heated
cavity was solved using the Radial Basis Function-generated Finite Difference (RBF-FD) method~\cite{tolstykh2003rbffd}, with a static, manually prescribed refined node distribution. While that approach produced accurate results, the
refinement pattern had to be chosen a priori. In this paper we extend that
framework with an automatic $h$-adaptive discretisation procedure  based on a lightweight variability indicator that quantifies the dissimilarity of the approximated differential operator values within a single stencil. Starting
from a coarse initial layout, the node density is iteratively adjusted within a single simulation run, based on the local variability indicator,
concentrating nodes in boundary-layer regions where the shear-thinning
behaviour produces steep velocity and temperature gradients.

In \autoref{sec:methods} we introduce the meshless numerical method and outline the adaptivity procedure. In \autoref{sec:problem} we introduce the natural convection driven non-Newtonian fluid flow that we use as the test case. Finally, in \autoref{sec:results}, we examine the impact of parameters used in adaptivity and demonstrate that the adaptive approach achieves comparable accuracy at a fraction of the computational cost of a uniform fine discretisation.

\section{Methods}
\label{sec:methods}
\subsection{RBF-FD}
In meshless approximation the domain is discretised with scattered computational points $\vec{x}_i$ at which we store the values of the underlying fields. Node locations are selected with a dedicated advancing front algorithm~\cite{slak2019generation} that supports variable density based on internodal distance function $h(\vec{x})$.

We construct stencils $S_i$ with size $s=15$ that contain the indices of $s$ closest nodes to $\vec{x}_i$. Linear operators in computational nodes
\begin{equation}
	(\L u)_i \approx \sum_{j=1}^{s} w_{i, j} u_{S_i(j)},
	\label{eq:operatorApprox}
\end{equation}
are approximated as a scalar product between node-specific weights $w$ and field values $u$ in neighbouring nodes. We use the RBF-FD method~\cite{leborne2023guidelinesRBFFD} with the basis of 3rd order polyharmonic splines and 2nd order monomial augmentation to determine the approximation weights. Further details about the numerical implementation are available in~\cite{Rot2025}.

\subsection{Adaptive discretisation}
\label{sec:adaptivity}
This approach to adaptivity is based on the thesis that if the properties of the underlying field within a single approximation stencil differ too much the resulting accuracy is likely to be degraded. We introduce variability indicator 
\begin{equation}
	\delta_i^\L = \frac{ \max_{j \in S_i} | \L u_i - \L u_j |}{\avg_{j \in S_i} | \L u_j |} 
	\label{eq:indicator}
\end{equation}
that quantifies the discrepancy in values of operator $\L$ within a stencil. Normalisation with average value is introduced for the indicator to remain relevant across a wide range of values $\L u_i$ at the cost of problematic areas where $\avg_{j \in S_i} | \L u_j | \approx 0$. Additional thresholding for the normalisation should be considered in future work. 

In this work we limit ourselves to the first derivatives and adapt based on
\begin{equation}
	\delta_i = \max\left(\delta_i^{\frac{\partial}{\partial x}}, \delta_i^{\frac{\partial}{\partial y}}\right),
\end{equation}
but it might be beneficial to incorporate all of the operators that appear in investigated equations.

The most important parameters for adaptivity are the refinement threshold $\refine$ and the derefinement threshold $\derefine$. When the value $\delta_i$ rises above the former we increase the local node density and decrease it when it falls below the latter. In practice, we construct a vector $\vec{h}$ of distances to the closest neighbouring node and multiply $h_i$ with the refinement factor $k=1.5$ when below $\derefine$ and divide with $k$ when above the refinement threshold. If exceeded we force the user-defined $h_\mathrm{min}$ and $h_\mathrm{max}$ constraints. Afterwards, smoothing is applied to $\vec{h}$ to keep the node density gradient $\|\grad h\|$ below $\Delta = 0.05$. Shepard interpolant~\cite{Gordon1978Shepard} with 3 neighbours is used to interpolate $\vec{h}$ to $h(\vec{x})$ that can be used in the advancing front algorithm~\cite{slak2019generation} to re-populate the domain with new density nodes. The values of $k$, $\Delta$, and Sheppard neighbours were selected based on intuition and require further research. An example of nodes marked for (de)refinement and the resulting changes in density for four consequent steps of adaptivity are shown in \autoref{fig:refineDerefineScatter}.


After we re-discretise the domain with new node density we need to re-interpolate the field values. This is again done with the Shepard interpolant, now with 10 neighbours. This could alternatively also be done with a RBF-FD interpolant or even-better with a divergence-free interpolant~\cite{drake2020divFreeRBF} for the velocity field.

\section{Test problem}
\label{sec:problem}

\begin{figure}
	\centering
	\includegraphics[width=\linewidth]{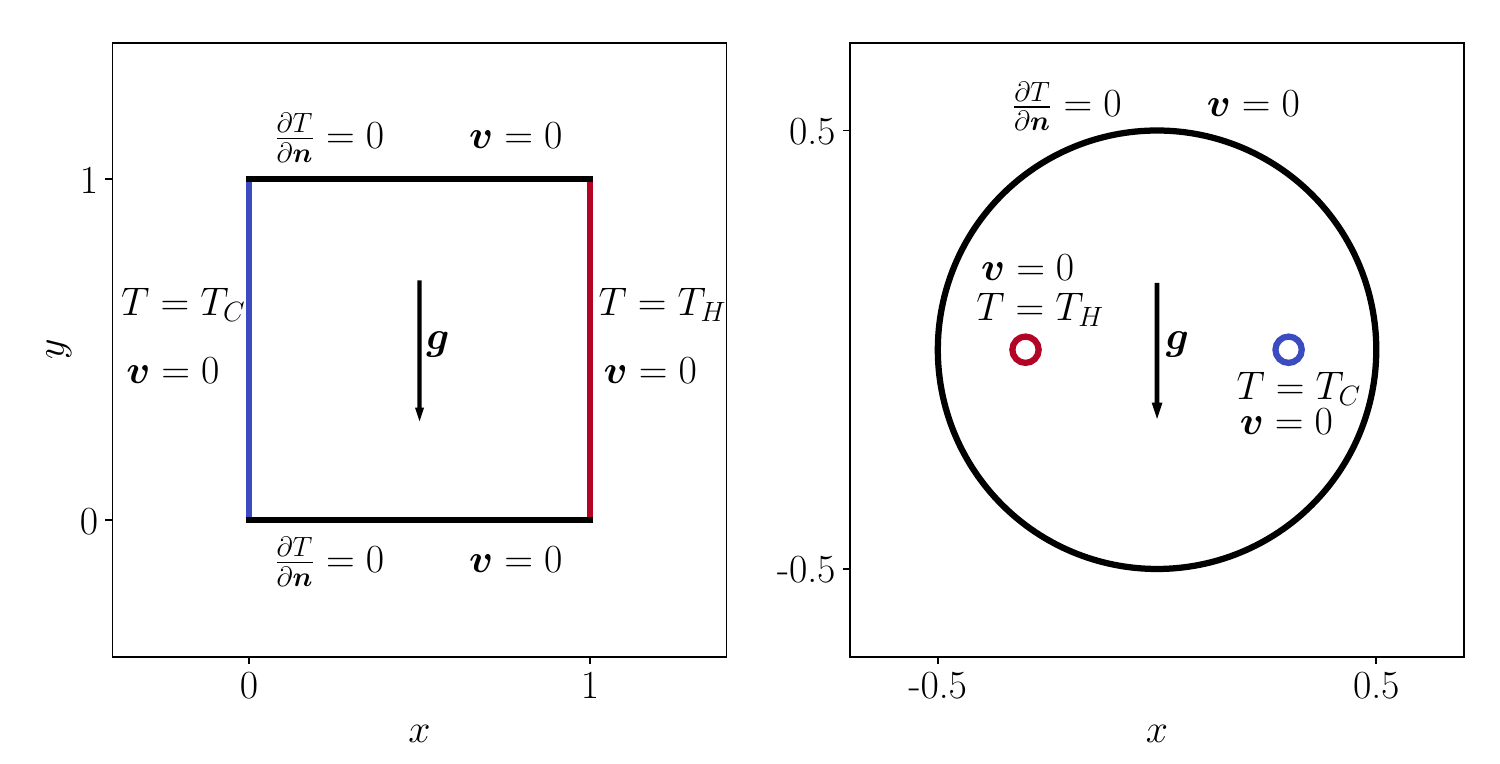}
	\caption{Sketch of the geometry and boundary conditions for the two considered differentially heated cavity cases. The de Vahl Davis case on the left and the spherical cavity case on the right. The hot boundaries are marked in red and the cold in blue.}
	\label{fig:caseSketch}
\end{figure}

\begin{figure}[b]
	\includegraphics[width=\linewidth]{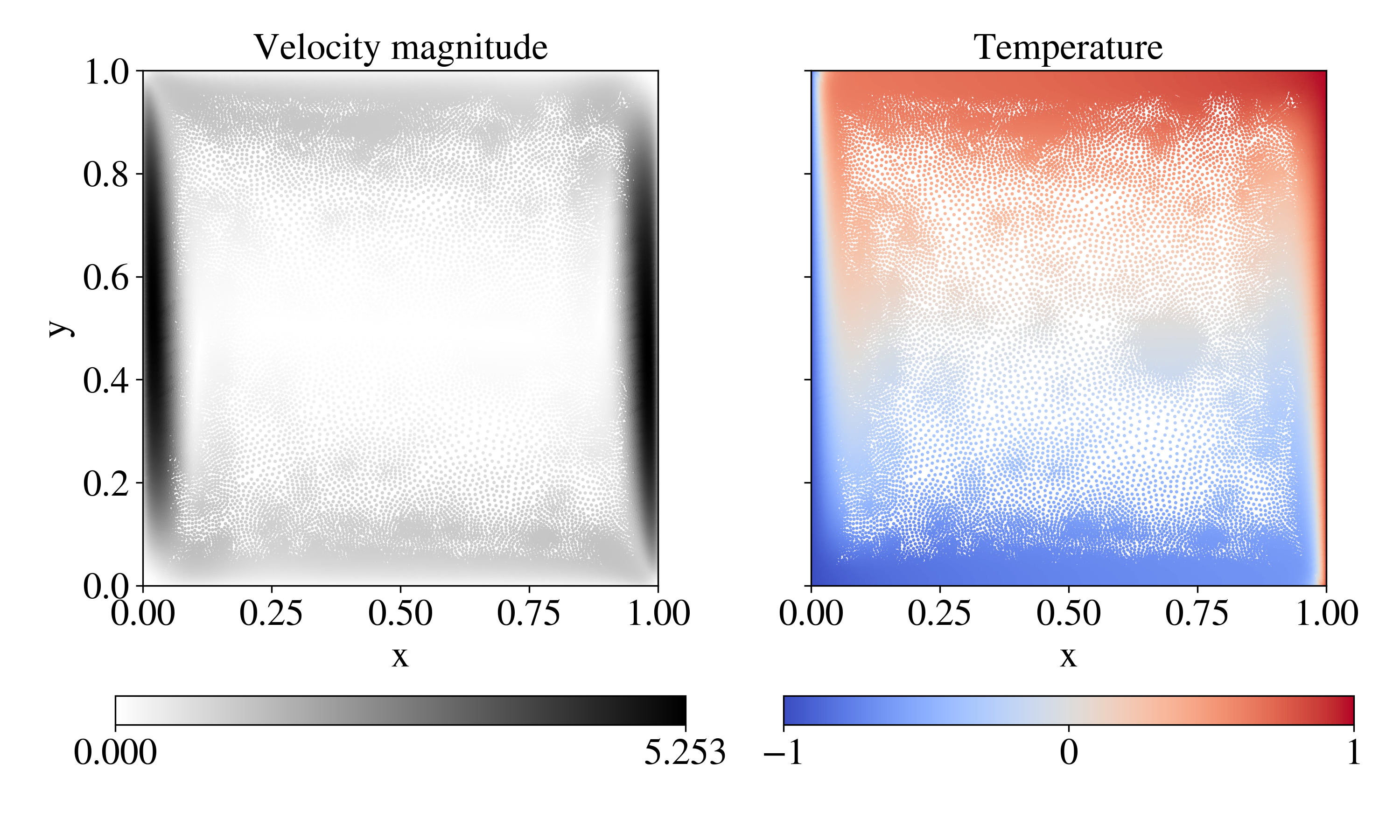}
	\caption{Velocity magnitude and temperature field for the de Vahl Davis case with $\mathrm{Ra} = 10^5$ and adaptive discretisation with $h_\mathrm{min} = 10^{-3}$, $\refine = 4$, and $\derefine = 1$ at $t=7$.}
	\label{fig:solutionFields_dvd}
\end{figure}

\begin{figure}[b]
	\includegraphics[width=\linewidth]{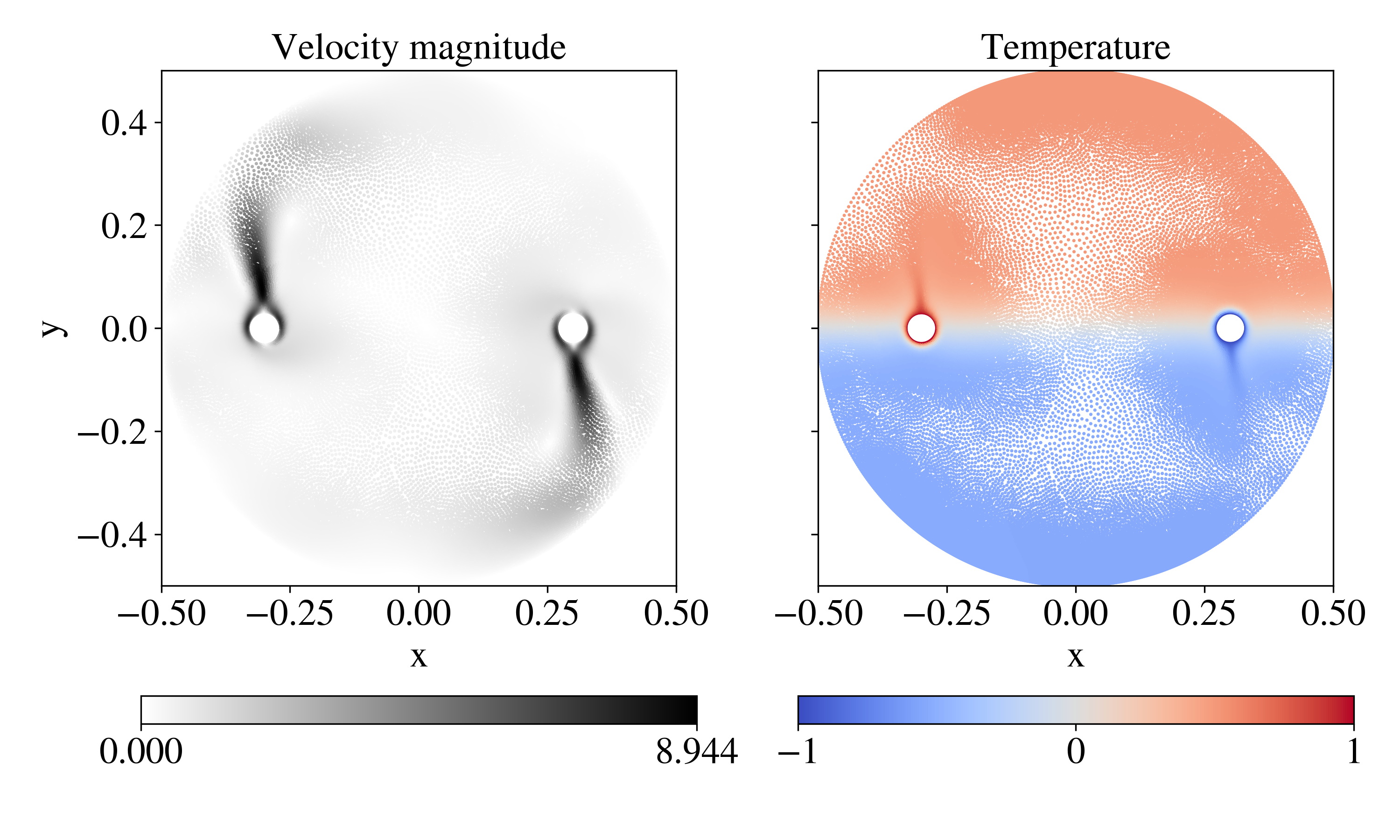}
	\caption{Velocity magnitude and temperature field for the Sphere case with $\mathrm{Ra} = 10^6$ and adaptive discretisation with $h_\mathrm{min} = 10^{-3}$, $\refine = 4$, and $\derefine = 1$ at $t=13$.}
	\label{fig:solutionFields_circles}
\end{figure}

We apply the adaptive discretisation algorithm to the natural convection driven incompressible flow of a non-Newtonian fluid in a closed differentially heated cavity. We consider the two cases shown in \autoref{fig:caseSketch}. The first, shown on the left, is the well known de Vahl Davis~\cite{deVahl1983} case of a square cavity with insulated top and bottom while the left wall set to a constant cold temperature $T_C = -1$ and the right wall to a constant hot temperature $T_H = 1$. Such configuration results in a circular flow shown in \autoref{fig:solutionFields_dvd} where the fluid rises at the hot wall and falls at the cold. This case was selected because of the plethora of available reference solutions. A second synthetic case was added to investigate if proposed adaptivity behaves similarly across different geometries. We chose a spherical cavity that contains differentially heated spherical obstructions. The hot one on the left held at a constant $T_H$ and the cold one on the right held at a constant $T_C$ again cause a circular flow pattern shown in \autoref{fig:solutionFields_circles}

The dynamics are governed by a system of partial differential equations describing the conservation of mass, momentum and energy 
\begin{align}
	\div \vec{v} &= 0, \label{eq:physics1}\\
	\rho \frac{Dv}{Dt} &= -\grad 
	p + 
	\div(\eta \left(\grad\vec{v} + (\grad{\vec{v}})^T\right)) -\vec{g} \rho \beta T_\Delta, \label{eq:physics2}\\
	\rho c_p \frac{DT}{Dt} &= \div(\lambda 
	\grad
	T),\label{eq:physics3}\\
	\eta &= \eta_0 \left(\frac{1}{2}\norm{\grad{\vec{v}} + 
		(\grad{\vec{v}})^T}
	\right) ^{\frac{n-1}{2}},\label{eq:physics4}
\end{align}
with $\vec{v}$, $T$, $p$, $\eta$, $\rho$, $\vec{g}$, $\beta$, $T_\Delta$, $c_p$, $\eta_0$, $n$, representing the flow velocity field, temperature field, pressure field, viscosity field, density, gravity, thermal expansion coefficient, temperature offset, heat capacity, viscosity constant and non-Newtonian exponent, respectively. Equations are abbreviated with the material derivative $\frac{Dw}{Dt} = \pdv{w}{t} + \vec{v} \cdot \grad{w}$. \autoref{eq:physics4} represents the Ostwald-de Waele power law model~\cite{yang2019comprehensive} for viscosity. The type and strength of the non-Newtonian behaviour is dictated by the exponent $n$. We set $n = 0.6$ resulting in a relatively strong shear-thinning.

Physical constants can be aggregated in two dimensionless numbers that characterise the flow behaviour. The Rayleigh number
\begin{equation}
	\mathrm{Ra}=\frac{\rho g \beta \Delta T L^{2 n + 1}}{\alpha^n \eta_0},
\end{equation}
where $\alpha = \frac{\lambda}{c_p \rho}$ is the thermal diffusivity, represents the ratio between buoyancy and thermal diffusivity while the Prandtl number
\begin{equation}
	\mathrm{Pr}=\frac{\eta_0}{\rho} \alpha^{n - 2} L^{2 - 2 n}
\end{equation}
is a material property that represents the ratio between momentum and heat diffusivity.  
We use $\mathrm{Pr} = 100$ for all presented results.

We use another dimensionless number as a convenient scalar observable for the behaviour of the dynamics. The Nusselt number
\begin{equation}
	\mathrm{Nu} = \frac{\grad T \cdot \hat{\vec{n}}}{\frac{T_H - T_C}{L}},
\end{equation}
where $\hat{\vec{n}}$ is the boundary normal, represents the ratio between the convective and the conductive heat transfer at the boundary. We use the average Nusselt value on the cold wall denoted as $\overline{\mathrm{Nu}}$.

Explicit solution procedure is preferred in this case due to the rapidly changing domain discretisation in the adaptive procedure. We solved the system of equations with the explicit Euler method and used the artificial compressibility method (ACM)~\cite{trojak2022acm} for the pressure-velocity coupling. Further details about the solution procedure are available in~\cite{Rot2025}.


\section{Results}
\label{sec:results}

The simulation is started with a constant node density with $h=0.0125$ and zero velocity \& temperature fields. We wait for the dynamics to develop before starting the adaptive discretisation at $t_\mathrm{start}$ and repeating every $\Delta t = 0.1$ until $t_\mathrm{end}$. Adaptivity is stopped prematurely if we reach the minimal prescribed inter-nodal distance $h_\mathrm{min}$ and changes in node count stagnate, which we define as $\frac{\Delta N}{N} < 0.02$. Simulation is continued after the last adaptivity step to reach a stationary state with the new node layout. We use $t_\mathrm{start}=3 \,\&\, t_\mathrm{end}=5$ for the de Vahl Davis case and $t_\mathrm{start}=9 \,\&\, t_\mathrm{end}=11$ for the sphere case. Maximum internodal distance $h_{\mathrm{max}} = 0.05$ is used in all cases. Final node density and solution fields are  shown in \autoref{fig:solutionFields_dvd} and \autoref{fig:solutionFields_circles}.

\begin{figure}
	\centering
	\includegraphics[width=\linewidth]{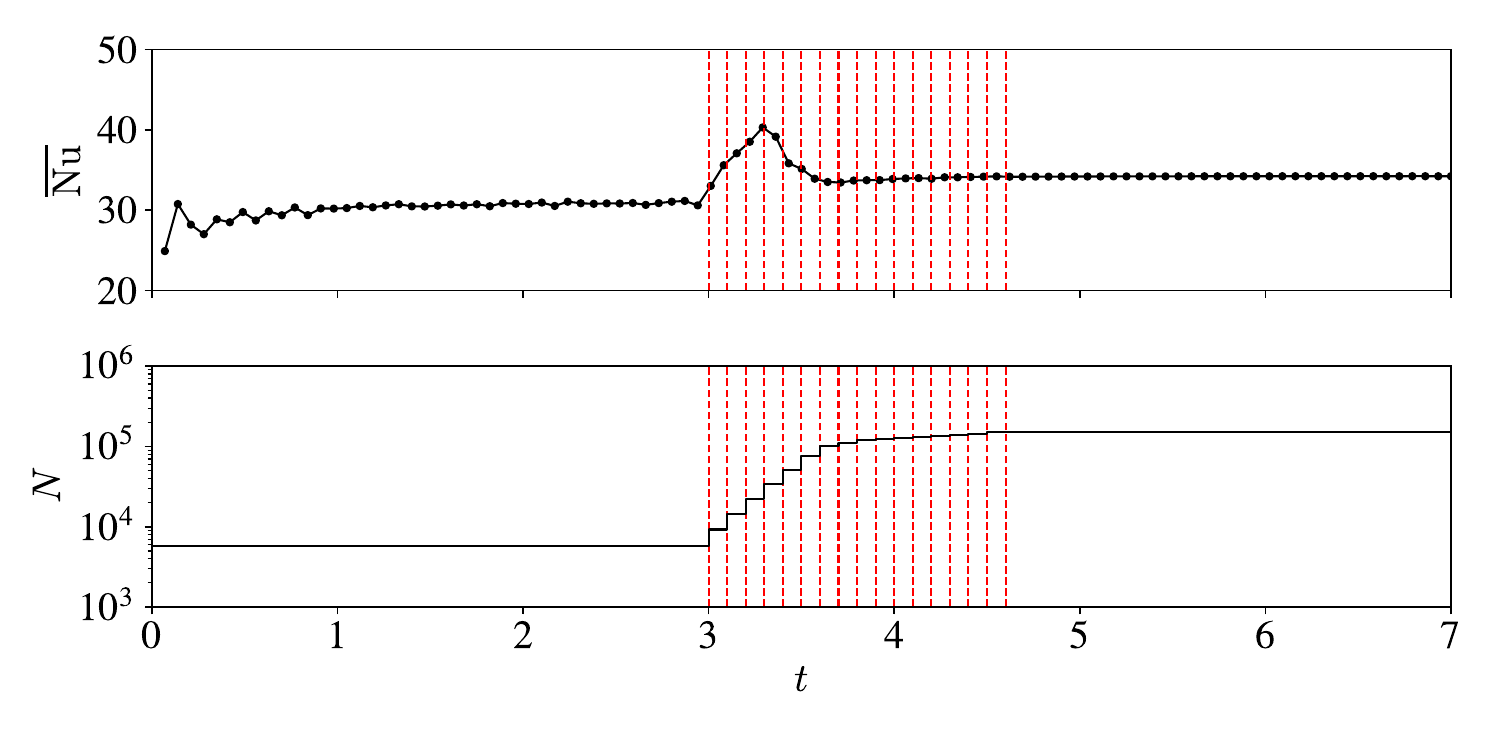}
	\caption{Evolution of system observables for the de Vahl Davis case with $\mathrm{Ra} = 10^6$. The top graph shows how the average nusselt number $\overline{\mathrm{Nu}}$ and the bottom how the node count $N$ changes with time and successive adaptive refinement iterations, indicted with red vertical dashed lines.}
	\label{fig:nusseltEvolution}
\end{figure}

Progression of adaptivity iterations for the de Vahl Davis case is visualised in \autoref{fig:nusseltEvolution}, showing the changes in the average nusselt number $\overline{\mathrm{Nu}}$ and number of nodes $N$ through time. The system is first evolved in time, almost reaching a stationary state before adaptivity iterations begin at $t_\mathrm{start}=3$. Number of nodes quickly rises with the first couple of adaptivity iterations before stagnating towards the final configuration. Similarly, $\overline{\mathrm{Nu}}$ first rises as the number of nodes close to the boundary increases, leading to erroneous result, as the boundary layer is not yet discretised accurately enough for the correct velocity profile to appear. This is quickly resolved as node density further increases and values $\overline{\mathrm{Nu}}$ to the correct value. This behaviour can be better understood in conjunction with the $h_\mathrm{min}$ convergence from \autoref{fig:adaptiveConvergence}. 

There are two main issues with the presented adaptivity procedure. The first is that each re-interpolation to the new node set introduces additional divergence into the velocity field. This is not critical in the considered case but could prove problematic with a  higher frequency of adaptivity iterations or a more unstable flow regime. Divergence can be corrected by introducing additional iterations of ACM pressure-velocity coupling when required but a true solution most likely includes divergence-free interpolation~\cite{drake2020divFreeRBF}.

The second problem lies with the normalisation in Eq.~\eqref{eq:indicator} that causes the variability indicator to spuriously increase when the denominator is close to 0 in areas with no significant dynamics that would require refinement. This problem can be readily observed in \autoref{fig:refineDerefineScatter} where small areas away from the boundary are marked for refinement. Patches of increased density that needlessly increase node count are still apparent in the final discretisation shown in \autoref{fig:solutionFields_dvd}. This problem could be resolved by introducing thresholding for the denominator or an alternate approach to variability indicator altogether.

\subsection{Adaptivity parameters}

\begin{figure}
	\centering
	\includegraphics[width=\linewidth]{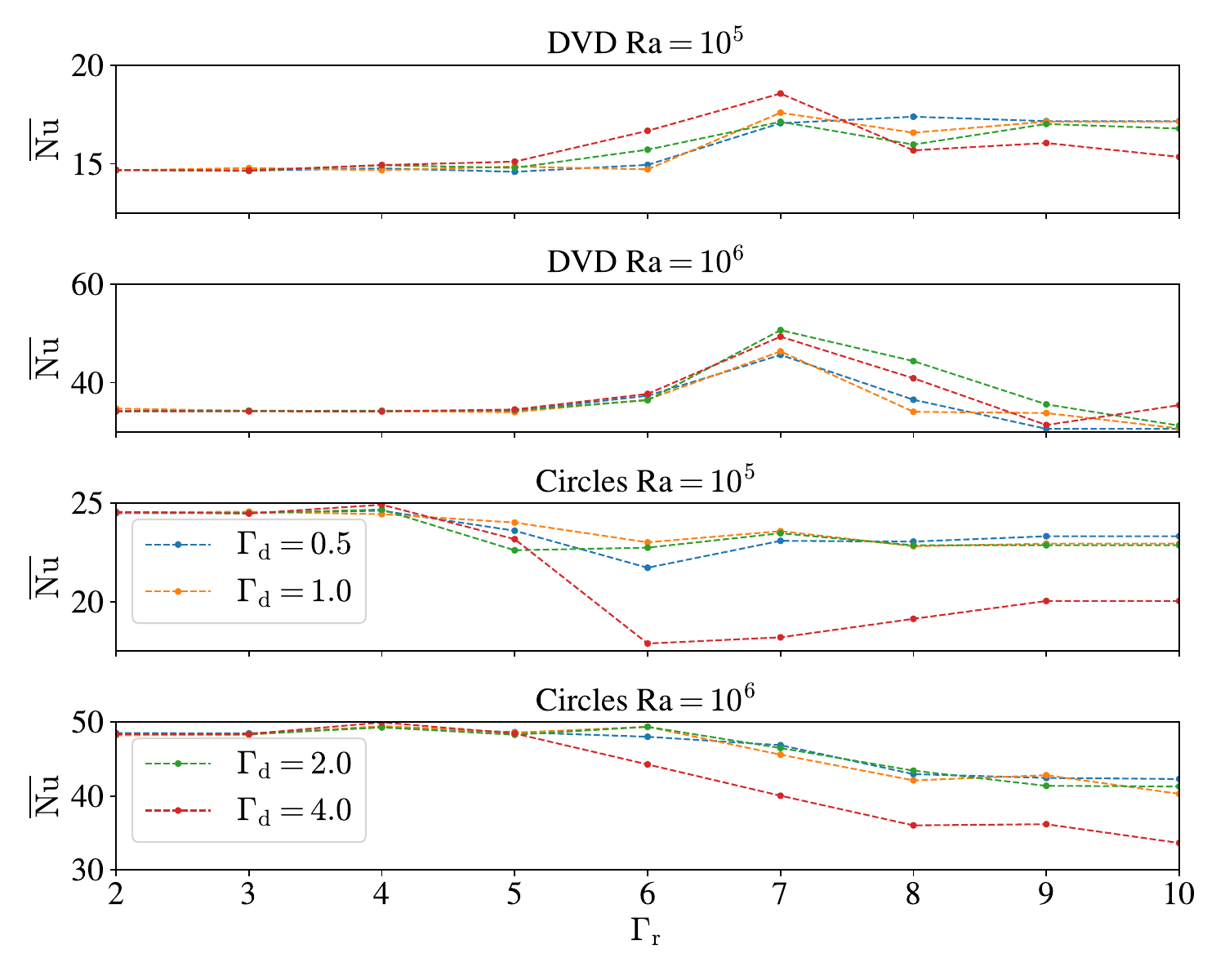}
	\caption{The average Nusselt number on the cold boundary $\overline{\mathrm{Nu}}$ as a function of refinement threshold $\refine$ for different values of derefinement threshold $\derefine$. Computed with $h_\mathrm{min} = 2 \cdot 10^{-3}$ for cases with $\mathrm{Ra} = 10^5$ and $h_\mathrm{min} = 10^{-3}$ for cases with $\mathrm{Ra} = 10^6$.}
	\label{fig:thresholdsNusselt}
\end{figure}

The two most important parameters for the adaptive procedure are the (de)refinement thresholds. Their selection represents a balancing act between accuracy and computational efficiency. A sweep across different (de)refinement threshold values and their impact on the average Nusselt number is shown in \autoref{fig:thresholdsNusselt}. We notice that the behaviour is similar for both of the considered $\mathrm{Ra}$ and both of the considered domains allowing us to discuss them together. In all cases, the value of $\overline{\mathrm{Nu}}$ stabilises as $\refine$ falls bellow $\sim 5$ prompting us to set $\refine = 4$ for all subsequent results.

\begin{figure}[b]
	\centering
	\includegraphics[width=\linewidth]{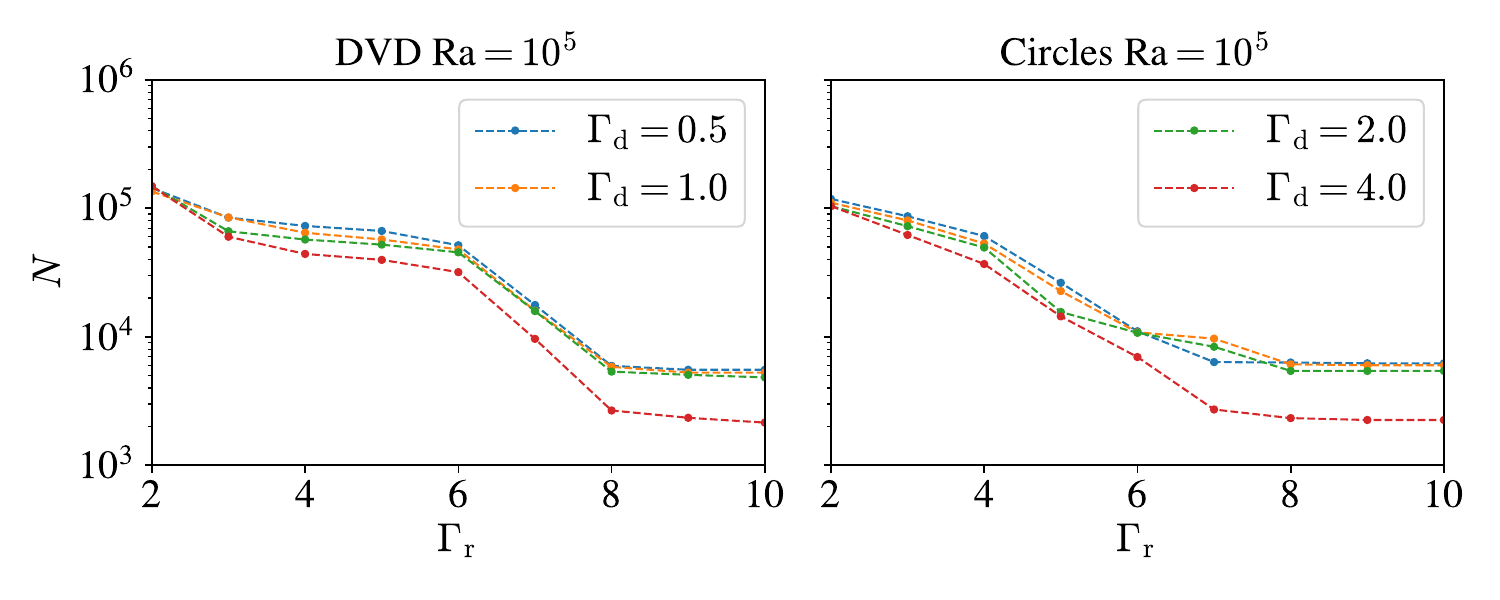}
	\caption{The final number of nodes in the discretisation as a function of refinement threshold $\refine$ for different values of derefinement threshold $\derefine$. Computed with $h_\mathrm{min} = 2 \cdot 10^{-3}$ for cases with $\mathrm{Ra} = 10^5$ and $h_\mathrm{min} = 10^{-3}$ for cases with $\mathrm{Ra} = 10^6$.}
	\label{fig:thresholdsCount}
\end{figure}

In terms of derefinement threshold, a significant difference only appears for $\derefine = 4$ and even there only at $\refine$ values that already return compromised results. The smaller considered $\derefine$ lead to comparable $\overline{\mathrm{Nu}}$. To better understand the computational complexity component of the threshold selection compromise we examine the impact of thresholds on the final number of nodes shown in \autoref{fig:thresholdsCount}. Aggressive derefinement does not lead to significant reduction in node count prompting us to use $\derefine = 1$ for the remainder of the results. Furthermore we can further justify the conservative selection of $\refine = 4$ as significant savings in node count only appear for $\refine > 6$ where \autoref{fig:thresholdsNusselt} already shows clear degradation in $\overline{\mathrm{Nu}}$.

The fact that the thresholds are relatively unaffected by the change in domain is encouraging in terms of applicability to an arbitrary geometry. It is unclear how they generalise to other equations and what the interplay with the approximation parameters is, warranting further investigation.   

\subsection{Convergence}

\begin{figure}[b]
	\centering
	\includegraphics[width=\linewidth]{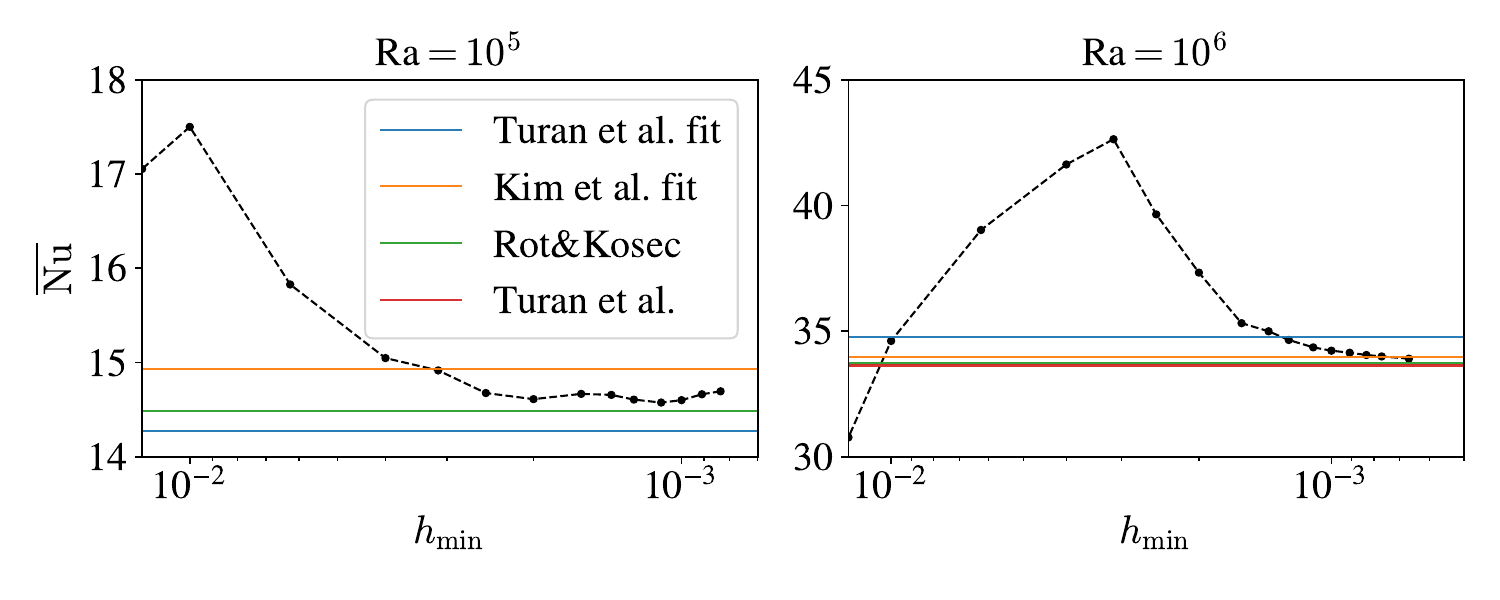}
	\includegraphics[width=\linewidth]{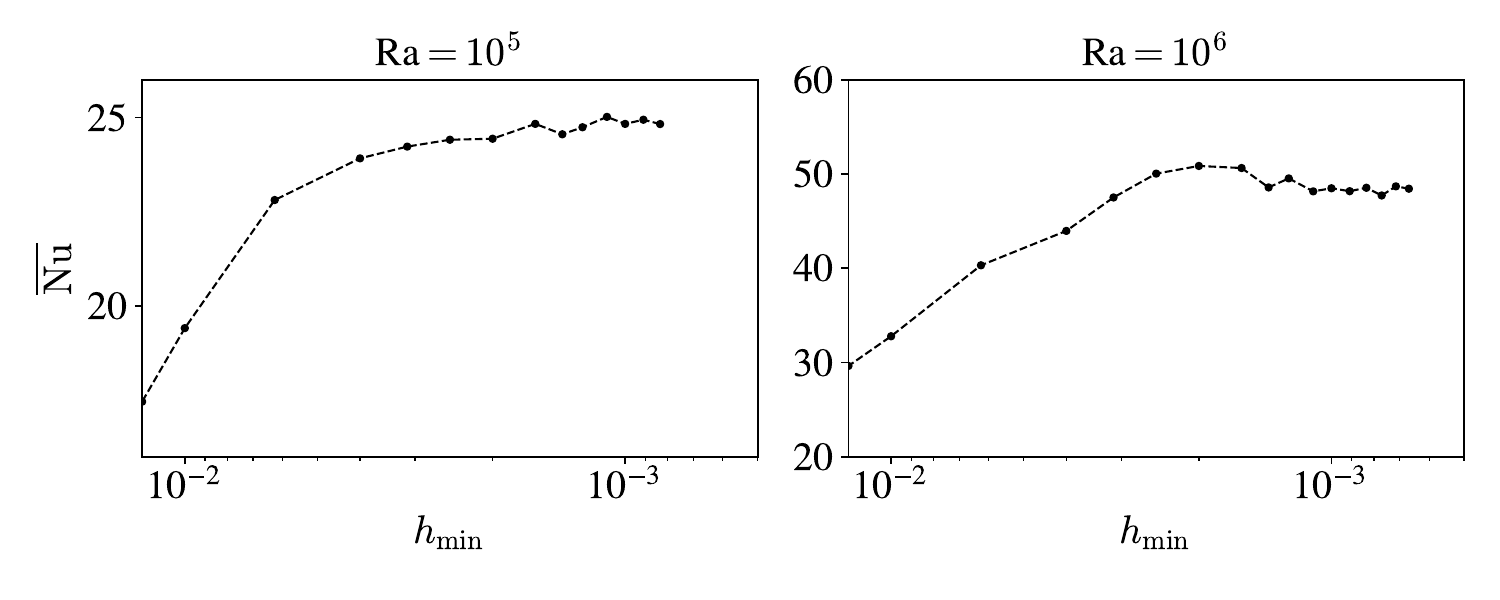}
	\caption{Convergence of the average Nusselt number on the cold boundary $\overline{\mathrm{Nu}}$ with decreasing minimal internodal spacing $h_\mathrm{min}$ for all of the considered cases. Results for the de Vahl Davis case, featured in the top row, are compared to reference solutions provided by Turan et al.\cite{turan2011laminar}, Kim et al. \cite{kim2003transient}, and the authors in their previous work\cite{Rot2025}. Results for the Sphere case are featured in the bottom row.}
	\label{fig:adaptiveConvergence}
\end{figure}

Now that we have established reasonable thresholds we can examine the consistency of the method and compare it with reference solutions. As the local discretisation accuracy increases in refined regions with ever-smaller minimal internodal distance $h_\mathrm{min}$ want the solution to converge to the same, ideally correct, solution fields. This is shown in terms of the scalar $\overline{\mathrm{Nu}}$ in \autoref{fig:adaptiveConvergence}. In the de Vahl Davis case, shown in the top row, the value settles between the existing reference solutions, close to the previous non-adaptive results obtained by the authors in their previous work~\cite{Rot2025}. We do not have a reference solution for the Sphere case but the $\overline{\mathrm{Nu}}$ value settles to a fixed value as we decrease $h_\mathrm{min}$, indicating consistent behaviour.

%

\subsection{Timing}
We continue to evaluate the adaptive procedure in terms of required computational time. Timing is performed on 4 cores of \texttt{Intel(R) Xeon(R) CPU E5520}. To improve the timing accuracy we disable the boost functionality and fix the frequency to 2.27 GHz.

\begin{figure}
	\includegraphics[width=\linewidth]{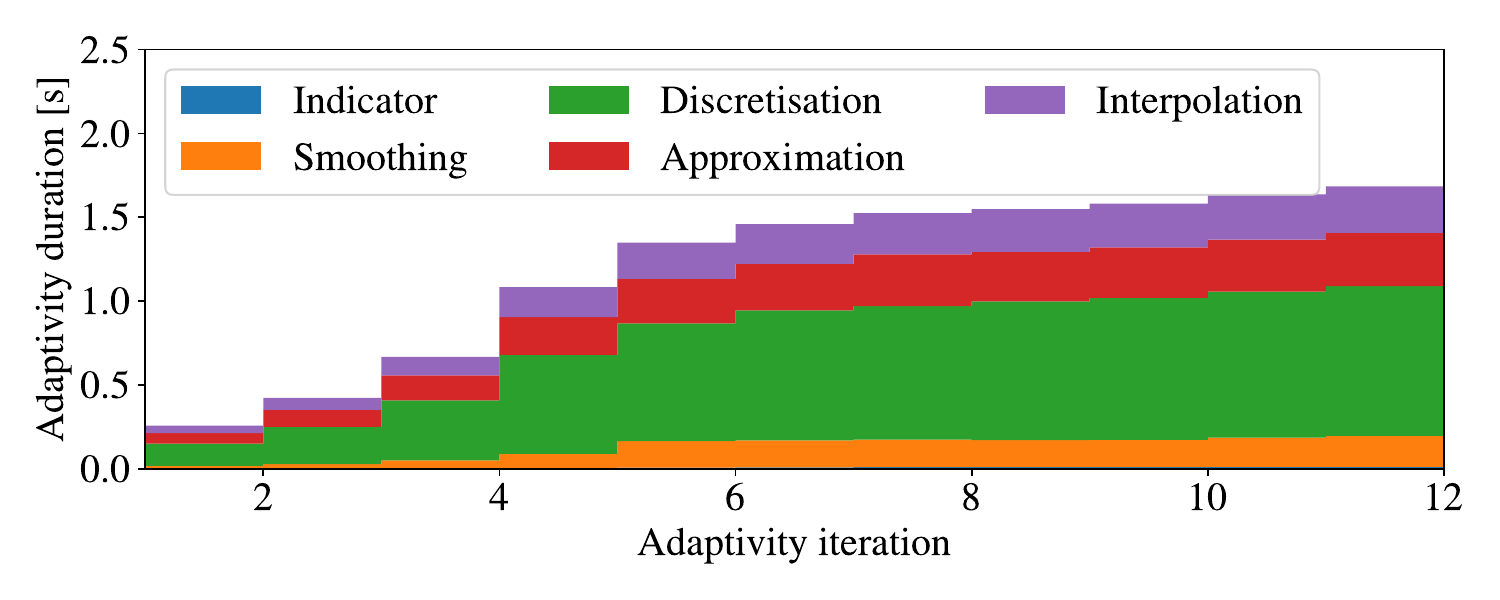}
	\caption{Computational cost of different parts of the adaptive refinement procedure for subsequent iterations of adaptivity. There are $N\approx10^4$ nodes in the discretisation before the first iteration and $N\approx5\cdot10^4$ before the last.}
	\label{fig:adaptPartTiming}
\end{figure}

We first compare the wall-time cost of individual parts of the adaptivity procedure shown in \autoref{fig:adaptPartTiming}. Relative cost of the parts remains relatively similar as domain size increases with subsequent adaptivity iterations. Re-discretising the domain with new nodes according to the new density proves to be the most expensive part of the procedure. Smoothing the density, calculating new approximation weights, and interpolating values from the old to the new node-set take approximately equal time. The cost of calculating the adaptivity indicator is practically negligible, suggesting that we could afford to use a more sophisticated approach. Note that the smoothing and the discretisation parts are performed in single thread mode and could benefit from parallelisation~\cite{Depolli2022ParallelFill}. 

Overall, the wall-time cost of one adaptive refinement is equal to $\sim120$ time evolution steps for the initial small domain before settling towards the equivalent of $\sim40$ steps when domain size increases.

\begin{figure}[b]
	\includegraphics[width=\linewidth]{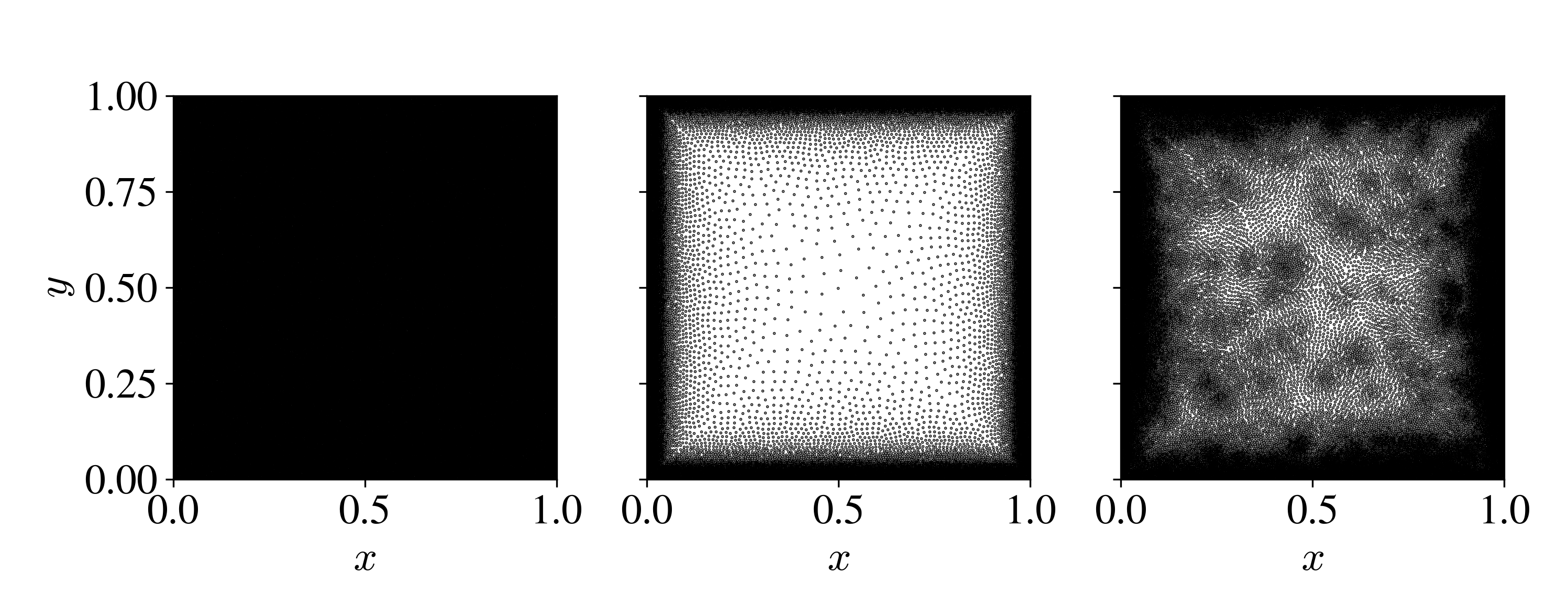}
	\caption{Comparison of the final node layouts for the three discretisation strategies we considered in the performance comparison. Constant density is shown on the left, the refined density in the middle and the automatically adaptive on the right.}
	\label{fig:discretisationNodeComparison}
\end{figure}

Finally, we compare the computational cost of the de Vahl Davis cavity flow computed with the adaptive refinement procedure to the static-in-time discretisation approaches considered in~\cite{Rot2025}. The comparisons are the constant density discretisation, the refined discretisation with a narrow band of high density close to the boundaries followed by a linear decrease towards the centre of the domain. The final node layouts for the considered discretisation strategies are shown in \autoref{fig:discretisationNodeComparison} with computational time and further details about the cases in \autoref{tab:timing}.
\begin{table}
	\begin{center}
		\begin{tabular}{|c|c|c|c|c|c|}
			\hline
			& $h_\mathrm{min}$                   & $h_\mathrm{max}$                   & $N$                     & $\overline{\mathrm{Nu}}$ & $t$[h]  \\
			\hline
			adaptive   & \multirow{3}{*}{0.0025} & 0.05   & 50809 & 39.0 & \bf{3.2} \\
			\cline{1-1} \cline{3-6}
			refined   &                         & 0.05   & 26961 & 39.8 & \bf{5.0} \\
			\cline{1-1} \cline{3-6}
			unrefined &                         & 0.0025 & 140134 & 38.8 & \bf{63.1} \\
			\hline
		\end{tabular}
	\end{center}
	\caption{A table showing differences in computational time between the refined, the unrefined and the adaptive discretisation for the DVD case with $\mathrm{Ra}=10^6$.}
	\label{tab:timing}
\end{table}

\begin{figure}
	\includegraphics[width=\linewidth]{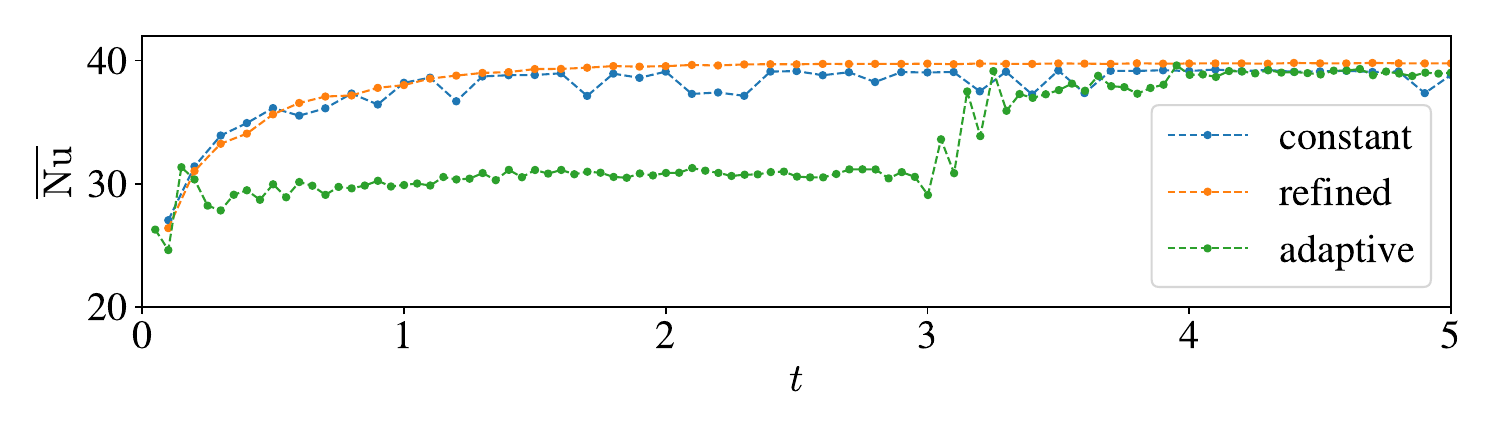}
	\caption{Evolution of $\overline{\mathrm{Nu}}$ throughout the solution procedure for the compared discretisation strategies.}
	\label{fig:discretisationNuComparison}
\end{figure}

We first observe that all of the proposed discretisations result in a comparable average Nusselt number, with the constant-in-time discretisation values straddling the adaptive. All of the considered discretisations have the same minimum internodal spacing $h_\mathrm{min} = 0.0025$\footnote{Note that while according to \autoref{fig:adaptiveConvergence} this selection of $h_\mathrm{min}$ does not yet result in an accurate $\overline{\mathrm{Nu}}$ value decreasing it further would lead to excessive computational time in the constant density case while not qualitatively impacting performance comparison.}. Constant density discretisation has by far the most nodes and an order of magnitude longer computational time. Refined procedure results in approximately half the nodes of the adaptive due to the aforementioned issues with spurious refinement in the central part of the domain that is readily apparent in \autoref{fig:discretisationNodeComparison}. The difference is not reflected in computational time though, with adaptive solution requiring $\sim 35\%$ less wall-time. Superior performance of the adaptive approach can be explained by examining the evolution of $\overline{\mathrm{Nu}}$ throughout the timing runs shown in \autoref{fig:discretisationNuComparison}. In the adaptive discretisation case the initial flow formation is computed with a sparse and computationally cheap node layout before starting adaptivity at $t_\mathrm{start}=3$ to identify the parts of the domain where increased density is required. The difference would be even more pronounced if we were not as conservative with the length of iteration after the last refinement step.

\section{Conclusions}

In this paper we outlined a procedure for dynamically adapting the density of computational nodes during the simulation and applied it to a fluid dynamics problem. The results show that the parametrisation for adaptivity is transferable between geometries and that we can exceed the computational efficiency of manually refined discretisation without requiring any \textit{a priori} knowledge about the problematic parts of the domain.

While the results show promise, further work is required both on testing the procedure on additional flow and non-flow problems, and investigating how including other operators impacts the variability indicator. Further computational efficiency gains are possible by resolving the thresholding issues that cause refinement in stagnant areas and developing better strategies for determining when to apply the adaptive iteration.


\bibliographystyle{IEEEtran}
\bibliography{adaptiveNN}

\end{document}